\documentstyle[amssym, onecolumn]{mn}
\input{epsf.tex}

\newcommand{\mincir}{\raise -2.truept\hbox{\rlap{\hbox{$\sim$}}\raise5.truept
\hbox{$<$}\ }}
\newcommand{\magcir}{\raise -2.truept\hbox{\rla669p{\hbox{$\sim$}}\raise5.truept
\hbox{$>$}\ }}
\newcommand{\minmag}{\raise-2.truept\hbox{\rlap{\hbox{$<$}}\raise 6.truept\hbox
{$>$}\ }}
\newcommand{\be}{\begin{equation}}
\newcommand{\ee}{\end{equation}}
\newcommand{\ba}{\begin{eqnarray}}
\newcommand{\ea}{\end{eqnarray}}
\newcommand{\brr}{\begin{array}}
\newcommand{\err}{\end{array}}
\newcommand{\bc}{\begin{center}}
\newcommand{\ec}{\end{center}}

\newcommand{\etal}{{et al.}~}

\newcommand{\p}{\partial}
\newcommand{\f}{\frac}

\newcommand{\w}{\omega}
\newcommand{\de}{\delta}

\newcommand{\al}{\alpha}

\newcommand{\fvphi}{\tilde{\varphi}}

\newcommand{\bfp}{{\bf p}}

\newcommand{\bfq}{{\bf q}}

\newcommand{\bfL}{{\bf L}}

\newcommand{\vphi}{\varphi}

\newcommand{\calD}{{\cal D}}

\newcommand{\calS}{{\cal S}}
\newcommand{\calT}{{\cal T}}
\newcommand{\calV}{{\cal V}}
\newcommand{\calF}{{\cal F}}	
\newcommand{\lan}{\langle}
\newcommand{\ran}{\rangle}

\newcommand{\frakC}{{\frak C}}
\newcommand{\frakT}{{\frak T}}

\begin{document}

\title{Correlations of Cosmic Tidal Fields}

\author[P. Catelan and C. Porciani]{Paolo Catelan$^1$ and Cristiano 
Porciani$^{2}$\\
    $^1$ California Institute of Technology, Mail Code 130-33, Pasadena,
     CA 91125, USA \\
    $^2$ Institute of Astronomy, Madingley Road, Cambridge CB3 0HA, UK }

\maketitle

\begin{abstract}
We study correlations amongst tidal fields originated by the
large--scale distribution of matter in the Universe. The two--point
tidal correlation is described as a rank-4 tensor, whose elements can
be written in terms of four fundamental scalar functions ranging, with
respect the spatial separation, from purely transversal to purely
longitudinal correlations. Tidal fields, both on galaxy and cluster
scales, reveal to be correlated over distances larger than the
mass--density correlation lenght, though traceless tidal fields show
anti-correlation between diagonal terms along orthogonal directions.
The cross-correlation between mass and tidal field is also analyzed.
These results are relevant for galaxy formation and the interpretation
of large scale weak lensing phenomena. \\

\noindent{\bf Key words:} galaxies: statistics -- large-scale
structure of Universe.
%\vspace{1cm}
\end{abstract}

\maketitle

\section{Introduction}
The role played by cosmic tidal fields in the evolution of mass
density fluctuations in the Universe has been recognized since the
very first studies of gravitational instability dynamics (see
e.g. Peebles 1980 for a historical perspective). Torques induced by
tidal fields presumably prompted the process of angular momentum
acquisition by a proto-galactic halo (Hoyle 1949; White 1984; Heavens
\& Peacock 1988; Catelan \& Theuns 1996). Tidal fields influenced the
dynamics of the collapse around primordial dark matter fluctuations
(Doroshkevich 1970; van de Weygaert \& Babul 1994; van de Weygaert \&
Bertschinger 1996; Bond \& Myers 1996). Consequently, they also
determined the mass and spatial distributions of virialized halos
(Monaco 1995; Catelan \etal 1998; Lee \& Shandarin 1998; Jing 1998,
1999; Porciani, Catelan \& Lacey 1999; Sheth \& Tormen 1999).

Only recently, however, the importance of {\it spatial correlations}
amongst cosmic tidal fields has been evidenced.  For instance, our
ability of reconstructing maps of the cosmic mass distribution from
weak gravitational lensing measurements (Bacon, Refregier \& Ellis
2000; Kaiser, Wilson \& Luppino 2000; van Waerbeke \etal 2000; Wittman
\etal 2000) is hampered by our ignorance about the distribution and
alignment of galaxy shapes.  As a first guess, one can assume that
shapes and orientations of background galaxies are uncorrelated, and
that the detected correlations amongst galaxy ellipticities are solely
due to the foreground mass distribution.  However, in the last few
months, a series of observations and numerical simulations evidenced
that both galaxy shapes (Croft \& Metzler 2000; Brown et al. 2000) and
angular momenta are intrinsically correlated (Heavens, Refregier \&
Heymans 2000; Pen, Lee \& Seljak 2000). Theoretical models for shape
alignments, whether induced by spin correlations or not, ascribe them
to spatial correlations of the underlying tidal field (Catelan,
Kamionkowski \& Blandford 2000; Crittenden \etal 2000).

In this Letter we analyze two-point statistics of the cosmic tidal
fields, addressing some fundamental issues that have been only
marginally discussed in the aforementioned literature. We determine,
for example, the size of a tidally correlated region, and the
characteristic length of the cross-correlation between linear density
fluctuations and tidal fields.  If intrinsic alignments of galaxy
shapes are ultimately determined by the cosmic tidal fields, these are
the fundamental scales involved in the process. In \S 2, we briefly
remind the definition of the cosmic tidal field.  In \S 3 we compute
the rank-4 tidal-tidal correlation tensor. Section 4 discusses
astrophysical implications and contains our conclusions.

\section{Cosmic tidal fields}

Let us consider the gravitational potential $\vphi(\bfq)$ originated
at the comoving Lagrangian position $\bfq$ by the primordial mass
density fluctuations $\de$ in the Universe.  We define $\vphi$ through
the Poisson equation, $\nabla^2\vphi = \de$, where the density $\de$
can be thought as the primordial density field extrapolated to the
present time.  For our purposes, it is irrelevant whether $\vphi$ is a
Gaussian field or not. The tidal field $\calD_{\al\beta}$ is defined
as a rank-2 symmetric tensor
\be
\calD_{\al\beta}\equiv \p_\al \p_\beta\vphi = 
-\int \f{d\bfp}{(2\pi)^3}\,p_\al\,p_\beta\,\fvphi(\bfp)\,
{\rm e}^{i\bfp\cdot\bfq}\;,
\label{1}
\ee
where $\p_\al \equiv \p/\p q_\al$, $\,\bfp$ is the comoving Lagrangian
wavevector and $\fvphi$ is the Fourier transform of $\vphi$.  More
rigorously, $\calD_{\al\beta}$ is known as deformation tensor; the
{\it tidal} field is properly the traceless rank-2 tensor defined in
terms of the gravitational potential $\vphi$ as
$\calF_{\al\beta}=[\p_\al\p_\beta -(1/3)\de_{\al\beta}\nabla^2]\vphi$.
In a similar fashion, it is possible to introduce a third tensor
describing the velocity {\it shear}.  In fact, the flow of an
irrotational fluid as the cosmic dust can be described by introducing
a peculiar velocity potential, $\vphi_v$.  The shear tensor is then
defined as $[\p_\al\p_\beta -(1/3)\de_{\al\beta}\nabla^2]\vphi_v$.
During the linear regime we consider in this paper,
%$\vphi_g$ and $\vphi_v$ coincide, 
$\vphi_v=\vphi$, and we are left with only one tensor
$\p_\al\p_\beta\vphi$ which can be called, indifferently, tidal or
shear tensor, if we are not interested exclusively in its traceless
part.

\section{Correlations amongst tidal fields}

Let us now compute explicitly the tidal-tidal correlation, which turns
out to be a rank-4 symmetric tensor:
\be
\frakC_{\al\beta\gamma\sigma}(\bfq)
\equiv
\lan\calD_{\al\beta}(\bfq_1)\calD_{\gamma\sigma}(\bfq_2)\ran = 
\p_\al\,\p_\beta\,\p_\gamma\,\p_\sigma \,\xi_\vphi(q)\;,
\label{2}
\ee
where $q\equiv |\bfq_1-\bfq_2|$, and $\xi_\vphi$ is the gravitational
potential correlation function,
$
\xi_\vphi(q) \equiv \lan\vphi(\bfq_1)\vphi(\bfq_2)\ran =  
(2\pi)^{-3}\int d\bfp\,P_\vphi(p)\,{\rm e}^{i\bfp\cdot\bfq}\;.
\label{3}
$
Here, $P_\vphi(p)$ is the gravitational potential power spectrum,
$\lan\fvphi(\bfp)\fvphi(\bfp')\ran \equiv (2\pi)^3\de_D(\bfp +
\bfp')P_\vphi(p)$, and $\de_D$ denotes the Dirac function.  The
explicit calculation of the right hand side of eq.(\ref{2}) is
somewhat toilsome; the reader should bear in mind that $\p_\al =
(dq/dq_\al)(d/dq)= (q_\al/q) d/dq$, then apply iteratively this
identity to $\xi_\vphi$ and its radial derivatives. At the end of the
day, it turns out that the tidal-tidal correlation tensor can be
decomposed as follows
\be
\frakC_{\al\beta\gamma\sigma}(\bfq) =
T(q)\,\calT_{\al\beta\gamma\sigma} + 
V(q)\,\calV_{\al\beta\gamma\sigma} + 
S(q)\,\calS_{\al\beta\gamma\sigma} \;, 
\label{4}
\ee
where, respectively,
\be
T(q) \equiv \xi''''_\vphi(q) - \f{6}{q}\xi'''_\vphi(q) + 
\f{15}{q^2}\xi''_\vphi(q)
- \f{15}{q^3}\xi'_\vphi(q)\;, \ \ \ 
V(q) \equiv \f{1}{q}\xi'''_\vphi(q) - \f{3}{q^2}\xi''_\vphi(q) 
+ \f{3}{q^3}\xi'_\vphi(q)\;, \ \ \ 
S(q) \equiv \f{1}{q^2}\xi''_\vphi(q)- \f{1}{q^3}\xi'_\vphi(q)\;;
\label{5-7}
\ee
with $\xi'_\vphi \equiv d\xi_\vphi(q)/dq$ and so forth. In addition,
\be
\calT_{\al\beta\gamma\sigma} \equiv 
\f{q_\al q_\beta}{q^2}\f{q_\gamma q_\sigma}{q^2}
\label{8}
\ee
\be
\calV_{\al\beta\gamma\sigma} \equiv 
\f{q_\al q_\beta}{q^2}\de_{\gamma\sigma}+
\f{q_\al q_\gamma}{q^2}\de_{\beta\sigma}+
\f{q_\al q_\sigma}{q^2}\de_{\gamma\beta}+
\f{q_\gamma q_\sigma}{q^2}\de_{\al\beta}+
\f{q_\beta q_\sigma}{q^2}\de_{\gamma\al}+
\f{q_\beta q_\gamma}{q^2}\de_{\al\sigma}
\label{9}
\ee
\be
\calS_{\al\beta\gamma\sigma} \equiv 
\de_{\al\beta}\de_{\gamma\sigma} +
\de_{\gamma\al}\de_{\beta\sigma} +
\de_{\al\sigma}\de_{\gamma\beta}\;,
\label{10}
\ee
where $\de_{\al\beta}$ is the Kronecker symbol. 
\footnote{Note that, in linear perturbation theory, the tidal
correlation tensor evolves proportionally to $D(t)^2$, with $D(t)$ the
growth factor of density perturbations.}
Crittenden \etal (2000) report [their eq.(34)] an alternative
expression for eq.(\ref{4}).  The equivalence of the two results can
be shown using text-book relations like, for example, $\xi_\vphi''(q)
= -\xi_\de(q) - 2q^{-1}\xi_\vphi'(q)$, and $\,\xi_\vphi'(q)=
-q^{-1}\xi_\vphi(q) + (2\pi^2)^{-1}\int
dp\,p^{-1}P_\de(p)\,\cos(qp)/qp$, and $\xi_\vphi'''(q) = (q/5)J_5(q)$,
and $\xi_\vphi''''(q) = \xi_\de(q) -(4/5)J_5(q)$ and so forth. Here
$\xi_\de(q)\equiv \lan\de(\bfq_1)\de(\bfq_2)\ran$, and $J_n(q)\equiv n
q^{-n}\int_0^q d\w\,\w^{n-1}\,\xi_\de(\w)$.  Eventually, one gets:
$T=\xi_\de+(5/2) J_3-(7/2) J_5$, $V=(1/2) J_5-(1/2) J_3$, $S=(1/6)
J_3-(1/10) J_5$.

\begin{figure}
\centerline{
\epsfxsize= 8 cm \epsfbox{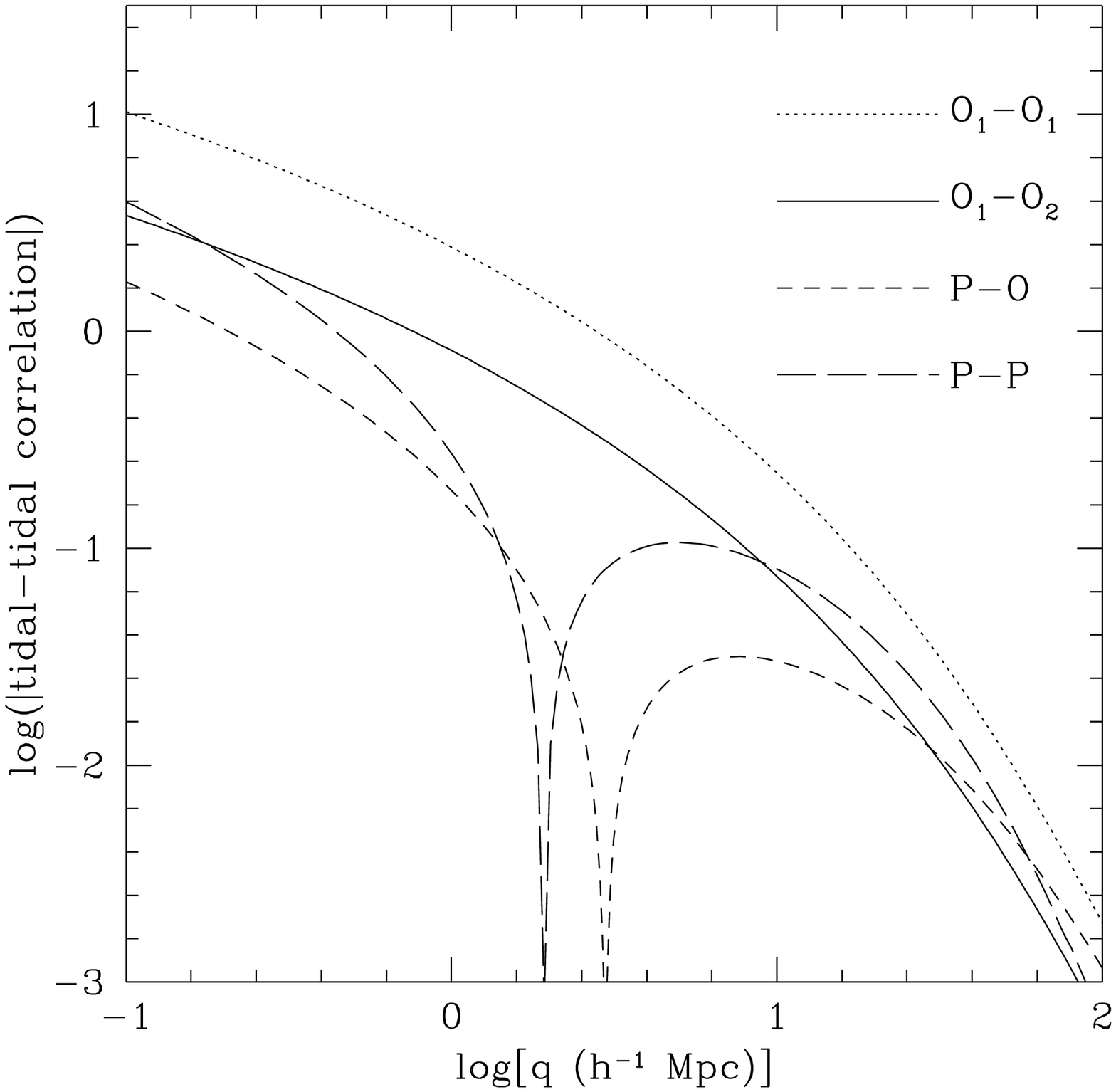} 
\epsfxsize= 8 cm \epsfbox{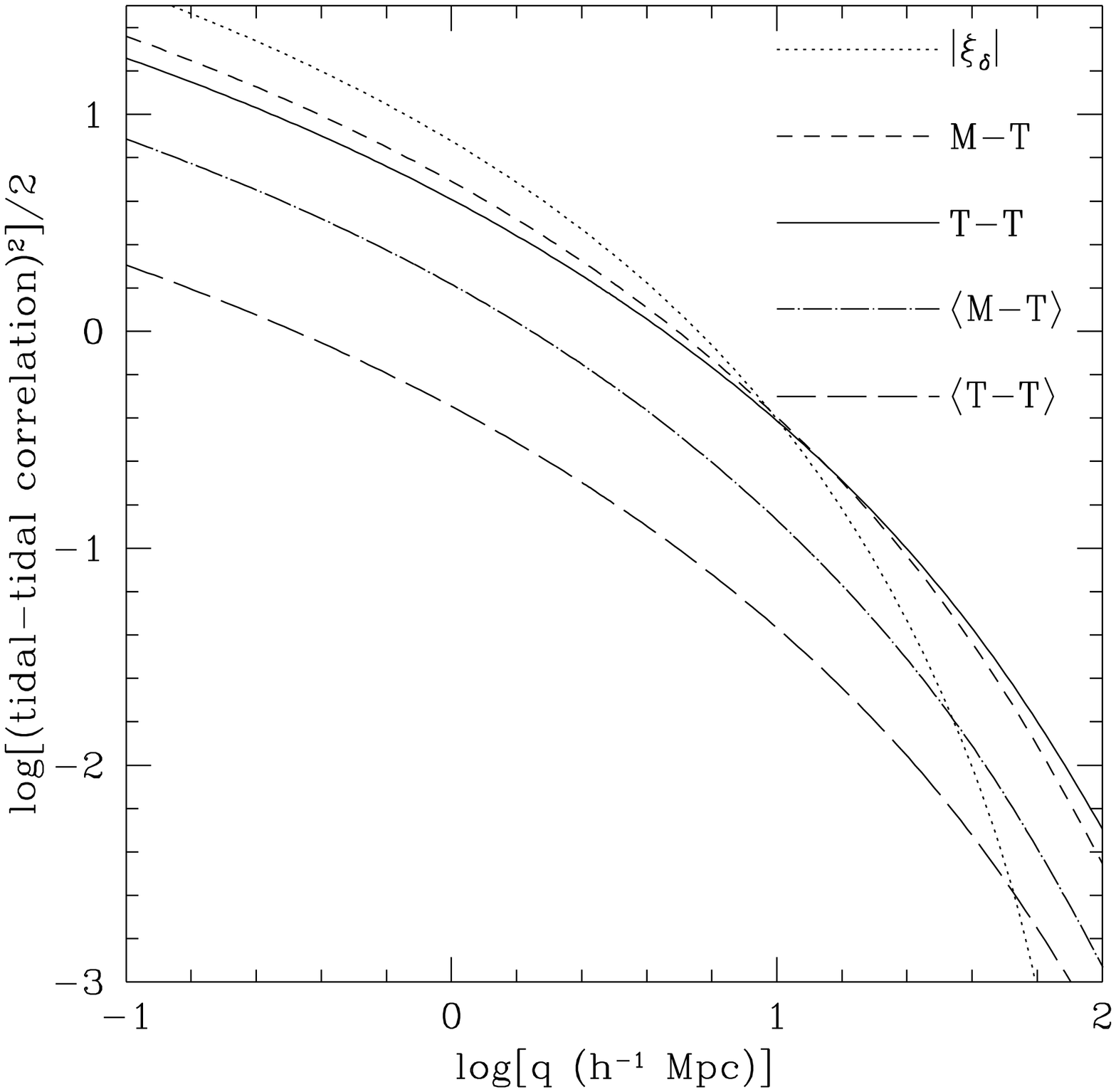}} 
\caption{Linear two-point correlations of the deformation tensor
vs. spatial separation, in a $\Lambda$CDM cosmology.  The separation
vector lies along the $z$-axis.  All the amplitudes are linearly
extrapolated to the present time.  {\it Left panel.} The solid line
shows the correlation between components along the two directions
perpendicular to the separation vector (O1-O2),
$\frakC_{1122}=\frakC_{1212}=\frakC_{2211}$; the dotted line shows the
correlation between diagonal components orthogonal to the separation
vector (O1-O1), $\frakC_{1111}=\frakC_{2222}$.  The short-dashed line
shows the correlation $\frakC_{1133}=
\frakC_{1313}=\frakC_{3311}=\frakC_{2233}$ etc. (P-O).  The
correlation of the diagonal component along the separation is plotted
with a long-dashed line (P-P).  {\it Right panel.} The correlation
scalar $(\frakC_{\al\beta\gamma\sigma}
\frakC_{\al\beta\gamma\sigma})^{1/2}$ (solid line) is compared to
$|\xi_\delta|$ (dotted line), and to the mass-tidal correlation
$(\frakC_{\al\beta\gamma\gamma} \frakC_{\al\beta\sigma\sigma})^{1/2}$
(short dashed line).  The typical tidal-tidal correlation,
($\frakC_{\al\beta\gamma\sigma}
\frakC_{\al\beta\gamma\sigma}/81)^{1/2}$, (long-dashed line) and
mass-tidal correlation, $(\frakC_{\al\beta\gamma\gamma}
\frakC_{\al\beta\sigma\sigma}/9)^{1/2}$, (dot-dashed line) are also
plotted.}
\label{fig1}
\end{figure}

The most striking feature appearing in eq.(\ref{4}) is the anisotropy
of the tidal-tidal correlation function: the spatial derivatives
combine in different ways depending on the orientation of the
separation vector with respect to the frame of reference.  The
anisotropic structure of $\frakC_{\al\beta\gamma\sigma}$ correlations
can be evidenced by explicitly writing its 81 components.  For
simplicity, we choose the reference frame in such a way that the
positive $z$-axis lies along the separation vector
$\bfq=\bfq_1-\bfq_2$.  In this way, the only non-vanishing components
are
\be
\frakC_{1111}(q)=\frakC_{2222}(q)= 3\,S(q) = \f{1}{2}J_3(q)-\f{3}{10}J_5(q)\;,
\label{10b}
\ee
\be
\frakC_{1122}(q)=\frakC_{2211}(q)=\frakC_{1212}(q)= S(q)=\f{1}{6}J_3(q)-\f{1}{10}J_5(q) \;,
\label{10c}
\ee
\be
\frakC_{1133}(q)=\frakC_{2233}(q)=\frakC_{1313}(q)= \frakC_{2323}(q)=
S(q)+V(q)=-\f{1}{3}J_3(q)-\f{2}{5}J_5(q)\;,
\label{10c}
\ee
\be
\frakC_{3333}(q) = 3S(q)+6V(q)+T(q)= \xi_\de(q)-\f{4}{5}J_5(q)\;,
\label{10d}
\ee
(the components of the tensor are invariant under permutations of the
indices). In total, there are 21 non-zero elements in the whole
tensor.  With respect to the separation-axis, we could refer
eq.(\ref{10b}) to eq.(\ref{10d}) as the `orthogonal 1--orthogonal 1',
`orthogonal 1--orthogonal 2', `parallel--orthogonal' and
`parallel--parallel' tidal correlations: they can be considered as the
basic building blocks of the more general tensor in
eq.(\ref{4}). These fundamental functions are plotted in the left
panel of Fig. \ref{fig1} considering a $\Lambda$CDM cosmology.
\footnote{The model is specified by the following parameters.  Mass
density parameter $\Omega=0.3$, cosmological constant density
parameter $\Omega_\Lambda=0.7$, Hubble constant $H_0=100 \, h\, {\rm
km/s/Mpc}$, with $h=0.7$. A cold dark matter (CDM) power spectrum of
density fluctuations with primordial spectral index $n=1$ is assumed.
This power spectrum is normalized by requiring that rms density
fluctuation in a $8 \,h^{-1}$ Mpc sphere is $\sigma_8=0.9$ when
linearly extrapolated to present-day.}

Recently, the problem of estimating the size of the tidally correlated
{\it sphere} about any given point has become of relevance. To this
purpose, one has somehow to consistently get rid of the anisotropies
of the tidal correlation tensor.  The simplest way to attain this is
to average the correlation tensor over all the directions, $\langle
\frakC_{\al\beta\gamma\sigma} \rangle_{4 \pi} = [(1/15) T+(2/3) V+ S]
\,\calS_{\al\beta\gamma\sigma}= (1/15)\,
\xi_\de\,\calS_{\al\beta\gamma\sigma}$, and the contributions from the
off-diagonal elements cancel out.  Alternatively, one can saturate the
free indices in $\frakC_{\al\beta\gamma\sigma}(\bfq)$ in order to
construct {\it scalar} (i.e. rotationally invariant) quantities.
Obviously, there is not a unique way to do this. A first attempt might
be to consider the simplest $\frakC_{\al\al\al\al}=\sum
\p_\al^4\xi_\vphi$; unfortunately, this operation also eliminates
purely tidal contributions from cross-terms like
$\p_1^2\p_2^2\xi_\vphi$ and so forth. The next simplest scalar one
might look at is $\frakC_{\al\al\beta\beta}$; but, trivially, this is
the mass-mass correlation function, deprived, again, of any
information about the actual tidal correlations (induced by the off
diagonal elements of the tidal tensor). One is forced to consider
higher-order scalars. The simplest of them is obtained saturating
$\frakC_{\al\beta\gamma\sigma}$ with itself:
$\frakC_{\al\beta\gamma\sigma}\frakC_{\al\beta\gamma\sigma}$. [Note
that this quantity will be quadratic in the underlying correlations,
so one might prefer instead to consider the scalar
$(\frakC_{\al\beta\gamma\sigma}\frakC_{\al\beta\gamma\sigma})^{1/2}$.]
Notably, this is the same quantity to which Lee \& Pen (2000) and Pen,
Lee and Seljak (2000) ascribe the leading-order spin-spin
correlations.

Summarizing, we assume here that the scalar
$(\frakC_{\al\beta\gamma\sigma}\frakC_{\al\beta\gamma\sigma}/81)^{1/2}$
is an estimate of the typical tidal correlation about any given point
in the Universe. We find, after saturating the indices,
\be
\frakC_{\al\beta\gamma\sigma}(\bfq)\frakC_{\al\beta\gamma\sigma}(\bfq)=
\lan\calD_{\al\beta}(\bfq_1)\calD_{\gamma\sigma}(\bfq_2)\ran^2\,= 
\xi_\de(q)^2 -\f{8}{5}\,J_5(q)\,\xi_\de(q) + 
2\,J_3(q)^2 - 4 \,J_3(q)\,J_5(q) + \f{14}{5}\,
J_5(q)^2\;.
\label{11}
\ee
A discrepancy with the results reported by Lee \& Pen (2000) appears
here. Their eq.(H5), in fact, claims that
$\frakC_{\al\beta\gamma\sigma}^2 = \xi_\de^2$, where the several other
terms appearing in the right hand side of our eq.(\ref{11}), all of
the same order of magnitude, are neglected. (Rather, the correct
expression would be $\frakC_{\al\al\beta\beta}^2 = \xi_\de^2$.)

\begin{figure}
\centerline{
\epsfxsize= 8 cm \epsfbox{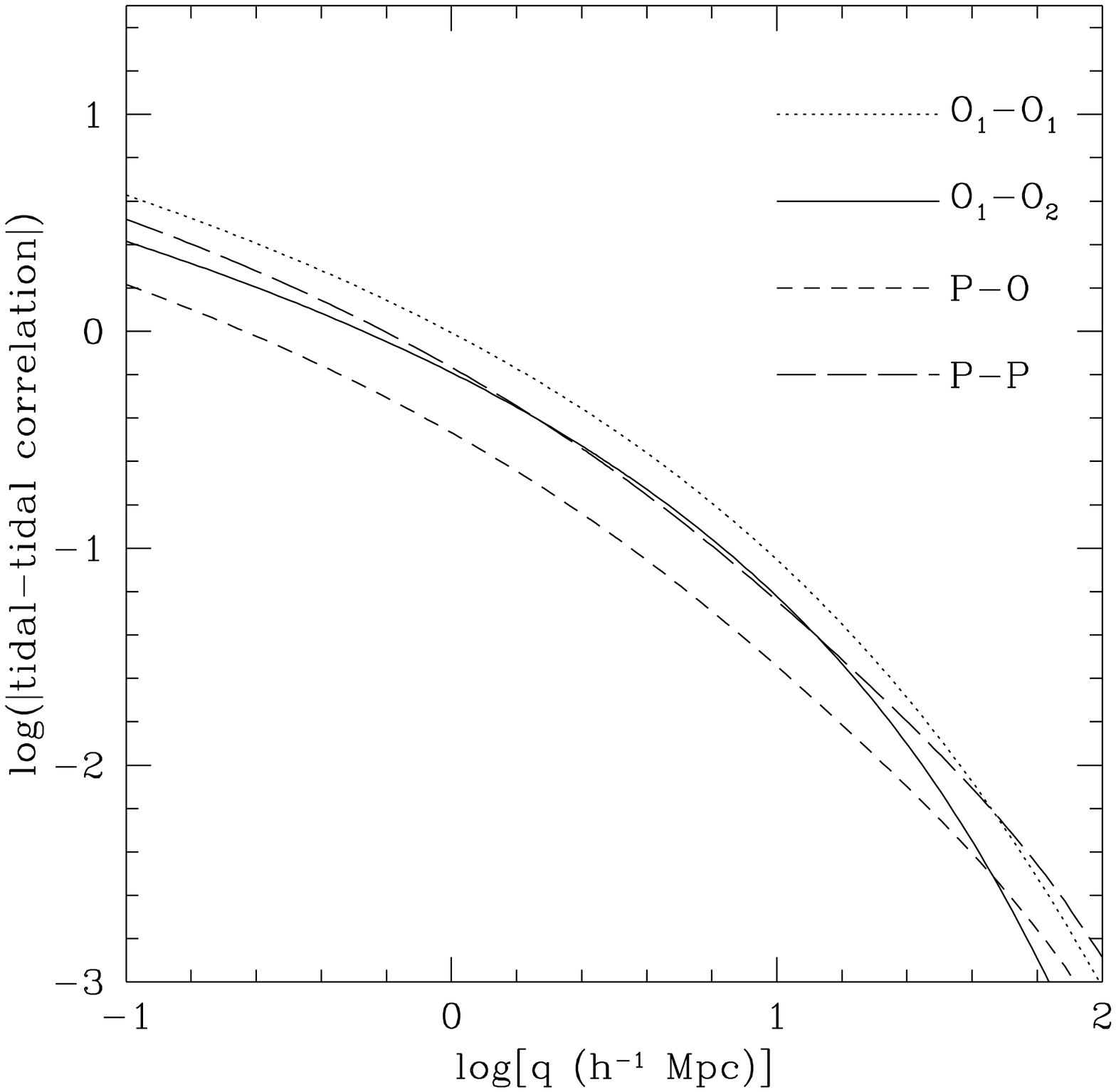} 
\epsfxsize= 8 cm \epsfbox{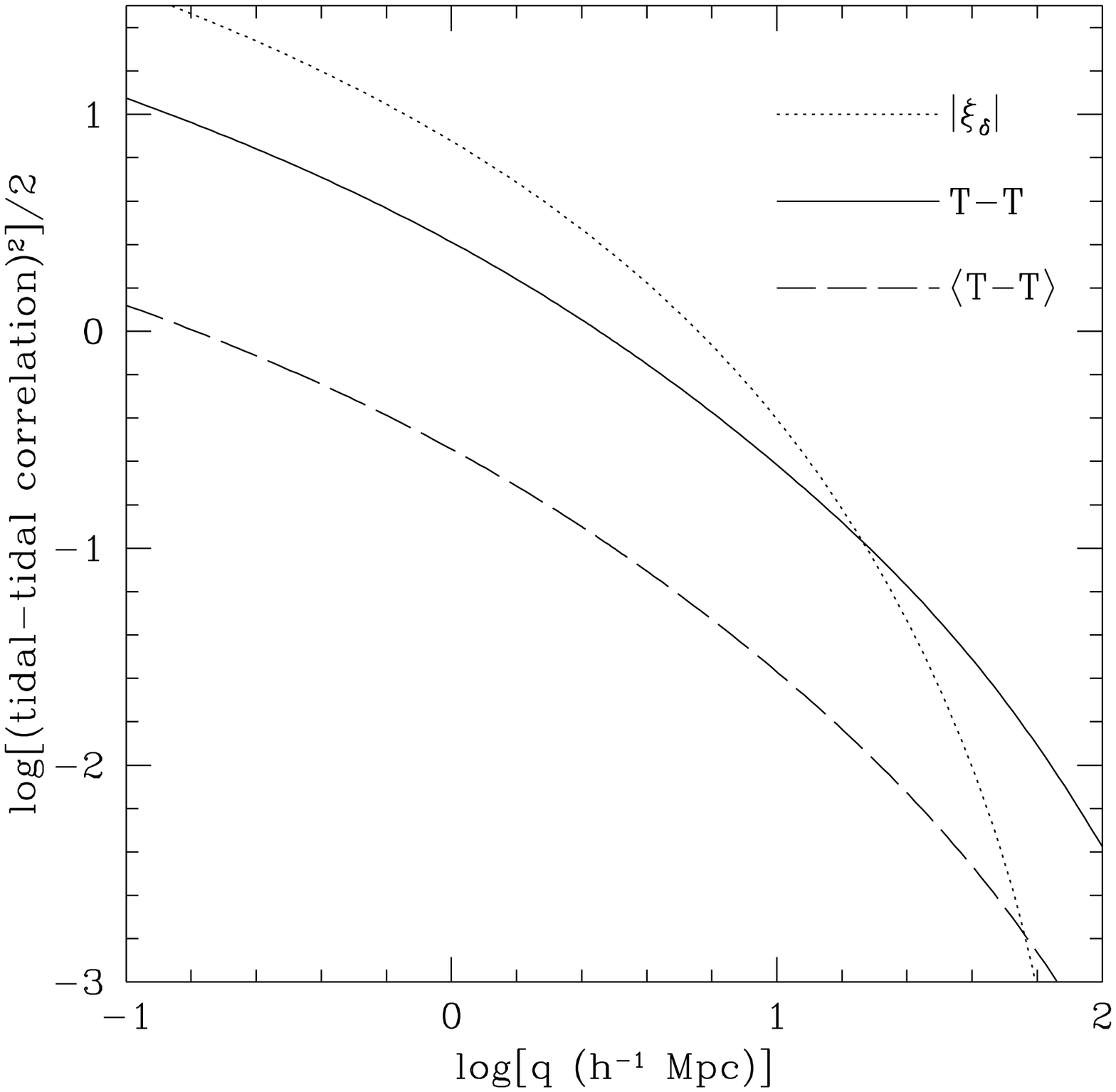}}
\caption{As in Fig. \ref{fig1}, but for the traceless tidal field
correlation $\frakT_{\al\beta\gamma\sigma}$.  {\it Left panel.}  All
symbols are like in Fig. (\ref{fig1}).  The solid line represents
$\frakT_{1122}=\frakT_{2211}$; $\frakT_{1212}$ has not been plotted
since it coincides with solid--line curve in Fig.\ref{fig1}.
Analogous considerations are valid for $\frakT_{1133}=\frakT_{2233}$
(short-dashed line) and $\frakT_{1313}=\frakT_{2323}$ (short-dashed
line in Fig. \ref{fig1}).  The functions $\frakT_{1122}$ and
$\frakT_{1133}$ are negative, indicating anti-correlation bewteen
diagonal components of the tidal tensor. {\it Right panel.}  The
scalars $(\frakT_{\al\beta\gamma\sigma}
\frakT_{\al\beta\gamma\sigma})^{1/2}$ (solid line) and
$(\frakT_{\al\beta\gamma\sigma}
\frakT_{\al\beta\gamma\sigma}/81)^{1/2}$ (long-dashed line) are
compared to $|\xi_\delta|$ (dotted line).}
\label{fig2}
\end{figure}

Another interesting quantity to study is the cross-correlation between
mass concentrations and tidal fields.  This is, in fact, a special
kind of tidal-tidal correlation, since the mass fluctuations
correspond to the trace of the tidal matrix.  Mass-tidal correlations
could be detected in the vicinity of galaxy clusters, e.g. through
weak gravitational lensing.
From the Poisson equation, the mass-tidal rank-2 correlation tensor
can be obtained from the more general rank-4 tensor
$\frakC_{\al\beta\gamma\sigma}$ via the saturation of two indices, for
example $\gamma$ and $\sigma$; we obtain
\be
\frakC_{\al\beta\gamma\gamma}(\bfq) =
\lan\calD_{\al\beta}(\bfq_1)\calD_{\gamma\gamma}(\bfq_2)\ran = 
[T(q) + 7 V(q)]\,\f{q_\alpha q_\beta}{q^2} + 
[V(q) + 5S(q)]\,\delta_{\alpha\beta} = 
[\xi_\de(q) - J_3(q)]\,\f{q_\alpha q_\beta}{q^2} + 
\f{1}{3}J_3(q)\,\delta_{\alpha\beta}\;.
\label{13}
\ee
Again, a rotationally invariant scalar can be constructed from
$\frakC_{\al\beta\gamma\gamma}(\bfq)$ by squaring it;
\be 
\frakC_{\al\beta\gamma\gamma}(\bfq)\frakC_{\al\beta\sigma\sigma}(\bfq) = 
\xi_\de(q)^2 -\f{4}{3}\,J_3(q)\,\xi_\de(q) + \f{2}{3}\,J_3(q)^2  .
\label{14}
\ee
This function is plotted in the right panel of Fig.\ref{fig1}. Note
that
$[\frakC_{\al\beta\gamma\gamma}(\bfq)\frakC_{\al\beta\sigma\sigma}
(\bfq)/9]^{1/2}$ is an estimate of the typical mass-tidal
cross-correlation.  It is also of interest to consider the correlation
of the traceless tidal field $\frakT_{\al\beta\gamma\sigma}(\bfq)=
\langle \calF_{\al\beta}(\bfq_1) \calF_{\gamma\sigma}(\bfq_2)
\rangle$:
\ba
\frakT_{\al\beta\gamma\sigma}(\bfq) &=& 
\frakC_{\al\beta\gamma\sigma}(\bfq)
-\f{1}{3} \delta_{\gamma\sigma}\,\frakC_{\al\beta\eta\eta}(\bfq)
-\f{1}{3} \delta_{\al\beta}\,\frakC_{\eta\eta\gamma \sigma}(\bfq)
+\f{1}{9} \delta_{\al\beta}\,\delta_{\gamma\sigma}\,
\frakC_{\nu\nu\eta\eta}(\bfq)       \nonumber \\
&=& \frakC_{\al\beta\gamma\sigma}(\bfq)
+\f{1}{3}[J_3(q)-\xi_\de(q)]
\Big( 
\f{q_\al q_\beta}{q^2}
\delta_{\gamma\sigma}  +
\f{q_\gamma q_\sigma}{q^2}\delta_{\alpha\beta} \Big)+ 
\f{1}{9}[\xi_\de(q)-2J_3(q)]\,\delta_{\al\beta}\,\delta_{\gamma\sigma}\;.
\ea
For $\bfq=(0,0,q)$, only the following components do not vanish,
\be
\frakT_{1111}(q)=\frakT_{2222}(q)= \f{1}{9}\xi_\de(q)+\f{5}{18}J_3(q)-
\f{3}{10}J_5(q)\,,
\label{tfa}
\ee
\be
\frakT_{1122}(q)=\frakT_{2211}(q)=\f{1}{9} \xi_\de(q)-\f{1}{18} J_3(q)-
\f{1}{10}J_5(q)\,,
\label{tfb}
\ee
\be
\frakT_{1133}(q)=\frakT_{2233}(q)=\f{2}{9}\xi_\de(q)-\f{2}{9}J_3(q)+
\f{2}{5}J_5(q)\,,
\label{tfd}
\ee
\be
\frakT_{3333}(q) = \f{4}{9} \xi_\de(q)+\f{4}{9} J_3(q)-\f{4}{5} J_5(q)\,,
\label{tfe}
\ee
and also $\frakT_{1212}(q)= \frakC_{1212}(q), \; \frakT_{1313}(q)=
\frakC_{1313}(q),\; \frakT_{2323}(q)=\frakC_{2323}(q)$.  The
corresponding correlation scalar is then
\be
\frakT_{\al\beta\gamma\sigma}(\bfq)\frakT_{\al\beta\gamma\sigma}(\bfq) = 
\f{4}{9}\xi_\de(q)^2+ 8\Big[\f{1}{9} J_3(q)-\f{1}{5} J_5(q)\Big]\,\xi_\de(q)+
\f{14}{9} J_3(q)^2-4 J_3(q)\,J_5(q)+\f{14}{5} J_5(q)^2\;.
\label{tfsca}
\ee
This quantity is plotted in the right panel of Fig. \ref{fig2}.
\section{Discussion and conclusions}
A tidal correlation length for (proto) dark matter haloes can be defined by
smoothing the fluctuation field on the relevant scales, and
determining the separations at which the normalized correlation
functions, e.g.
$\frakC_{\al\beta\gamma\sigma}(\bfq)/|\frakC_{\al\beta\gamma\sigma}(0)|$
(the covariance functions divided by the corresponding variances),
assume a chosen value $C$.  At zero separation, we obtain
%$\frakC_{1111}(0)=3\sigma^2/15$ and $\frakC_{1122}(0)=\sigma^2/15$ and
%so forth, where $\sigma^2$ is the variance of density fluctuations.  
$\frakC_{\al\beta\gamma\sigma}(0)=\sigma^2/15
\,\calS_{\al\beta\gamma\sigma}$, where $\sigma^2$ is the variance of
density fluctuations.  The correlation lengths $q_{1/2}$ and $q_{1/3}$
(corresponding to $C=1/2$ and $C=1/3$ respectively) are reported in
Table 1 for a smoothing radius of $1 \,h^{-1}$ Mpc (galaxy scales) and
of $10 \,h^{-1}$ Mpc (cluster scales).  Top-hat smoothing is adopted
in every case.  Note that, by definition, $q_{1/2}$ and $q_{1/3}$ keep
constant during the linear evolution of the density field. For a
$\Lambda$CDM cosmology, we found that, with respect to the separation,
purely orthogonal tidal correlations (O1-O1 and O1-O2) tend to be
stronger than the longitudinal tidal correlations (P-O and P-P); this
holds true for the traceless tidal field as well. The latter shows
systematic anti-correlations, being the O1-O2 and P-O correlations
negative on the relevant scales.  In addition, the tidally correlated
sphere around any given point, computed here via the scalar
$\frakC_{\al\beta\gamma\sigma}^2$, is bigger than the mass correlated
sphere.

Lee \& Pen (2000) and Pen, Lee \& Seljak (2000) used the expression
$\frakC_{\al\beta\gamma\sigma}^2 = \xi_\de^2$ to quantify the
leading-order three-dimensional correlation amongst galaxy spin
directions $\lan|{\hat \bfL}\cdot{\hat \bfL}'|^2\ran$ which, in turn,
would model correlations amongst intrinsic galaxy shape alignments: a
fundamental issue for reconstruction of dark matter maps from weak
gravitational lensing measurements. A more comprehensive discussion of
the implications of our eq.(\ref{11}) for the correlations amongst
galaxy angular momenta and for the reconstruction of dark matter maps
from weak gravitational lensing measurements is beyond the scope of
this Letter. However, if intrinsic alignments of galaxy shapes are
ultimately determined by the cosmic tidal fields, the scales reported
in Table 1 must be the fundamental scales involved in the process.

The results reported here do not assume that the gravitational
potential is Gaussian distributed, but if it is so, then our analysis
gives a complete description of the tidal-tidal statistics.

\begin{table*}
\begin{minipage}{110mm}
\caption{Correlation length (in $h^{-1}$ Mpc) of the tidal field, for
different smoothing scales, $R_{\rm TH}$ (same units). 
Negative numbers denote anti-correlation lengths. $\frakC^2 = 
\frakC_{\al\beta\gamma\sigma}\frakC_{\alpha\beta\gamma\sigma}$, M-T
(mass-tidal)=$\frakC_{\al\beta\gamma\gamma}\frakC_{\al\beta\sigma\sigma}$. 
A $\Lambda$CDM cosmology is assumed.}
\label{symbols}
\begin{tabular}
    {@{}ccccccccccc}
$R_{\rm TH}$ & Trace free &Type                
                    & O1-O1& O1-O2 & P-O   & P-P  & $\frakC^2$ & M-T  & 
$\xi_\de$ \\ 
 1  & N & $r_{1/2}$ & 3.0  & 3.0   & 1.1   & 1.4  & 2.1  & 2.0  & 1.8 \\ 
 1  & N & $r_{1/3}$ & 5.1  & 5.1   & 1.3   & 1.8  & 3.4  & 3.1  & 2.6 \\ 
 1  & Y & $r_{1/2}$ & 2.5  & -3.3  & -1.4  & 1.4  & 1.9  & -    & 1.8 \\
 1  & Y & $r_{1/3}$ & 4.0  & -5.6  & -2.0  & 2.0  & 3.1  & -    & 2.6 \\
10  & N & $r_{1/2}$ & 19.1 & 19.1  & 9.6   & 7.3  & 13.9 & 13.0 & 11.7 \\ 
10  & N & $r_{1/3}$ & 25.8 & 25.8  & 11.9  & 8.9  & 22.3 & 19.2 & 15.1 \\ 
10  & Y & $r_{1/2}$ & 15.4 & -24.1 & -9.1  & 9.1  & 12.6 & -    & 11.7 \\
10  & Y & $r_{1/3}$ & 22.4 & -30.0 & -11.6 & 11.6 & 18.4 & -    & 15.1 \\
\end{tabular}
\end{minipage}
\end{table*}

\medskip
CP acknowledges the support of the EC RTN network ``The Physics of the
Intergalactic Medium''. PC thanks Marc Kamionkowski and Rupert Croft
for discussions. This project has been partially funded by NSF
AST-0096023.

\end{document}